\documentclass[referee,useAMS,usenatbib,usegraphicx]{mn2e}

\newcommand{\nus}{\it NuSTAR}
\newcommand{\xmm}{\it XMM-Newton}

\def\1705{4U~1705--44}
\def\saxj{SAX~J1808.4--3658}

\usepackage{subfigure}
\bibpunct{(}{)}{;}{a}{}{,}

\title[{\nus} and {\xmm} broad--band spectrum of \saxj]
{{\nus} and {\xmm} broad--band spectrum of {\saxj} during its latest outburst in 2015}

\author[T. Di Salvo et al.]{T. Di Salvo$^{1}$\thanks{E-mail:tiziana.disalvo@unipa.it},
A. Sanna$^{2}$, L. Burderi$^{2}$, A. Papitto$^{3}$, R. Iaria$^{1}$, 
\vspace*{0.3cm}
\\ {\LARGE \textup{A.~F. Gambino$^{1}$, A. Riggio$^{2}$} }
\vspace*{0.1cm}
\\
$^1$Universit\`a degli Studi di Palermo, Dipartimento di Fisica e Chimica, 
via Archirafi 36 - 90123 Palermo, Italy\\
$^2$Universit\`a degli Studi di Cagliari, Dipartimento
di Fisica, SP Monserrato-Sestu, KM 0.7, 09042 Monserrato, Italy \\
$^{3}$INAF -- INAF - Osservatorio Astronomico di Roma, via di Frascati 33, 
I-00044 Monteporzio Catone, Roma, Italy   
  } 

\begin{document}

\date{}

\maketitle

\begin{abstract}
The first discovered accreting millisecond pulsar, \saxj, went into 
X-ray outburst in April 2015. We triggered a 100 ks {\xmm} ToO, taken at 
the peak of the outburst, and a 55 ks {\nus} ToO, performed four days apart.    
We report here the results of a detailed spectral analysis of both the 
{\xmm} and {\nus} spectra. While the {\xmm} spectrum appears much
softer than in previous observations, the {\nus} spectrum
confirms the results obtained with {\xmm} during the 2008 outburst. 
We find clear evidence of a broad iron line that we interpret as produced 
by reflection from the inner accretion disk. For the first 
time, we use a self-consistent reflection model to fit the reflection 
features in the {\nus} spectrum; in this case we find a statistically 
significant improvement of the fit with respect to a simple Gaussian 
or diskline model to fit the iron line, implying that the reflection
continuum is also significantly detected. 
Despite the differences evident between the {\xmm} and {\nus} spectra, 
the smearing best-fit parameters found for these spectra
are consistent with each other and are compatible with previous results.
In particular, we find an upper limit to the inner disk radius of $\sim 12~R_g$.
In all the cases, a high inclination angle ($>50^\circ$) of the system is 
required.
This inclination angle, combined with measurements of the radial velocity of
the optical companion, results in a low value for the neutron star mass 
($<0.8\,M_\odot$), a result that deserves further investigation. 

\end{abstract}


\section{Introduction}

An accreting millisecond pulsar (hereafter AMSP) is a neutron star (NS) 
accreting mass from a low mass companion star ($\le 1 M_\odot$) and rotating at 
millisecond periods, as it is witnessed by the coherent pulsations in its X-ray 
light curve. This phenomenon is relatively rare among Low Mass X-ray Binaries 
(LMXBs) and is caused by the NS magnetic field, which is 
strong enough (given the accretion rate) to effectively funnel the accreting 
matter onto the magnetic poles. Most of the AMSPs are transient X-ray sources 
with recurrence times between two and more than ten years, and their outbursts 
usually last from a week to two months at most. {\saxj} (J1808 hereafter) 
is the first discovered AMSP \citep{Wijnands.etal:98}. The observed X-ray 
coherent pulsations are a fundamental probe of its dynamical and orbital 
state. A 60 ks long {\xmm} observation of this source revealed a broad 
($FWHM \sim 2$ keV) emission line at the energy of iron fluorescence emission 
\citep[][P09 hereafter]{Papitto.etal:09}. This broad line has been confirmed by Suzaku, which 
observed the source the day after with compatible spectral parameters within 
the errors \citep{Cackett.etal:09}.

Such broadened features are ubiquitous among accreting compact objects. 
First discovered in Active Galactic Nuclei \citep[AGN, see e.g.][for a review]
{Fabian.etal:2000}, these were subsequently observed in Galactic X-ray binaries 
containing black holes \citep[e.g.][]{Miller.etal:2004} or NS \citep[e.g.][]
{Cackett.etal:10}. To explain the broadness of these features it is usually 
assumed that they originate from reprocessed emission of the accretion disc, 
illuminated by the primary Comptonized spectrum. 
Experimental results support the view that, in several cases, the accretion 
disk is truncated at few gravitational radii from the central mass. Here 
Doppler shifts and relativistic boosting in a fast rotating plasma, along 
with the gravitational red-shift caused by the strong gravitational field 
of the compact object,
distort the line profile asymmetrically, broadening its shape up to 1 keV. 
Their shape thus depends on the geometry of the reflecting region of the disc, 
on the Keplerian velocity in the disc, and on its ionization state. 
%
In this case the line parameters allow to derive several important physical 
features with unprecedented accuracy. We just mention here the inclination of 
the disk with respect to the line of sight, which, in some cases \citep[e.g.][]
{Di_Salvo.etal:09}, is derived with an accuracy of just few degrees, and, most 
important, the inner disk radius, which, in some sources, is determined within 
1-2 Rg (where $Rg = G M / c^2$ is the gravitational radius). For these systems 
the compelling discovery of a fast spinning (extreme Kerr) black hole has been 
claimed based on the fact that the inner disc radius derived from the fit 
of the iron line lies below 6 Rg, which is the last stable orbit for a 
non-rotating (Schwarzschild) black hole \citep[see e.g.][for a review]
{Fabian:2005}.
Moreover, iron lines play a particularly important role if observed in pulsars, 
as they carry information about the magnetospheric radius. This radius indicates 
where in the disk stresses exerted by the magnetic field start to remove angular 
momentum from the matter. According to accretion theory, it depends on the 
magnetic field strength, on the accretion rate, and on the details of the 
interaction between the magnetic field and the accreting matter. 

Studies of the iron line profile could be hence a very powerful diagnostic tool 
to investigate the behaviour of matter in extreme gravitational fields and to 
effectively constrain the compactness (mass to radius ratio) of NSs
or the spin parameter of black holes in X-ray binaries. 
However, some authors argued against the disk origin of broad iron lines,
addressing the asymmetric broadening of these lines as caused by Compton 
down–scattering in a outflowing wind (e.g.\ \citealt[][]{Titarchuk.etal:2009};
see, however, \citealt[][]{Cackett.Miller:13}) or to pile-up distortions in
CCD spectra (\citealt{Ng.etal:10}; see, however, \citealt{Miller.etal:10} for
extensive simulations of pile-up effects). 
Strong pile-up effects on the line profile can be excluded; in fact, it has been 
shown that there is a good agreement between spectral parameters derived from 
CCD-based spectra with those derived from gas-based spectrometers 
\citep[see e.g.][]{Cackett.etal:12, Egron.etal:13}. More recently, observations 
with {\nus}, which does not suffer from pile-up, have confirmed that iron line
profiles in most LMXBs appear broad and asymmetric \citep[see e.g.][and references
therein]{Miller.etal:13, Degenaar.etal:15, King.etal:16, Sleator.etal:16, 
Ludlam.etal:17}.

Nevertheless an asymmetric broad iron line may be the 
effect of either reflection from a Keplerian disc or Compton broadening and/or 
down–scattering. However, if the origin of this line is from disc 
reprocessing, one would also expect the presence of a spectral hump between 20 
and 40 keV due to Compton scattering of the primary spectrum by the disc. Indeed 
this reflection hump has been observed in the spectrum of some NS LMXBs 
\citep[e.g.][]{Barret.etal:00, Piraino.etal:99, Yoshida.etal:93, Fiocchi.etal:07, 
DiSalvo.etal:2015}, usually with reflection amplitudes (defined in terms 
of the solid angle $\Omega/2\pi$ subtended by the reflector as seen from the corona) 
lower than 0.3, indicating a spherical geometry of the illuminating corona. 
Therefore, the use of broad-band, moderately high energy resolution spectra,
together with the use of self-consistent reflection models able to simultaneously
fit the iron line profile and the related Compton hump, are fundamental 
in order to probe the consistency of the parameters of the whole reflection
component and the reliability of the disk parameters derived from the so-called 
Fe-line method \citep[see also][]{Matranga.etal:2017}. 

To date 22 AMSPs have been discovered since 1998 \citep[see][for reviews]
{Patruno.Watts:2012, DiSalvo.Campana:2018}, the last ones discovered in 2018 
(IGR J17379$-$3747, \citealt[][]{Sanna.etal:2018}, and IGR J17591$-$2342, 
\citealt[][]{Sanna.etal:2018b}),
and three transitional pulsars, including IGR J18245$-$2452, the only transitional
pulsars that went into an X-ray outburst \citep{Papitto.etal:2013a}. Spectral studies 
at high resolution are fundamental in order to characterise their emission during 
outbursts \citep[see e.g.][for a review]{Poutanen:2006}. Besides an energetically 
dominating Comptonized component, one or two soft components are often detected
if enough statistics and spectral resolution is guaranteed. These soft 
components are interpreted as the emission arising from the accretion disc 
and from the NS hot spots. 
One of the most frequently recurring AMSP is {\saxj} which goes into outburst 
more or less regularly every 2--3 years. The light curve shape is also very 
regular with outburst peak fluxes between 60 and 80 mCrab (2--10 keV), and 
a subsequent slow decay on a timescale of $10-15$ days until the source decays 
below 16 mCrab and enters a low luminosity flaring state.
Its spin frequency is constantly decreasing at a rate 
($\sim 5 \times 10^{-16}$ Hz/s) compatible with the one expected from dipole 
emission of a $\sim 10^8$ G rotating pulsar \citep{Hartman.etal:2009, Sanna.etal:2017b}.
While this is the most probable explanation for such a deceleration, it was also 
proposed that the NS spin-down may be due to the emission of continuous Gravitational 
Waves \citep{Bildsten:1998}. This source is also charactised by a puzzling fast 
orbital period evolution
\citep{DiSalvo.etal:2008, Hartman.etal:2009, Burderi.etal:2009, Patruno.etal:2012b,
Sanna.etal:2017b}.   
The time scale of this evolution is so short (few $\times 10^6$ yr) that a non 
conservative evolution \citep[e.g.\ the so-called radio-ejection model,][]
{Burderi.etal:2001} or large short-term angular momentum exchange between 
the mass donor and the orbit, caused by gravitational quadrupole coupling due 
to variations in the oblateness of the companion, are indicated as possible 
explanations  \citep[see][for a discussion]{Sanna.etal:2017b}. 
Such a conclusion might establish an evolutionary link between (at least some) 
AMSPs and the so-called Black Widow Pulsars. 
To be confirmed, this scenario needs more measures, as quasi–cyclic period 
variations are expected in binaries \citep{Arzoumanian.etal:1994}. 

{\saxj} was observed with XMM-Newton during its 2008 outburst (see P09 for details). 
During this 60~ks observation a broad iron line ($\sigma = 1.1 \pm 0.1$ keV) was detected 
at an energy of $\sim 6.4$~keV. Modelling this line according to the disc reflection 
hypothesis (diskline model) allowed P09 to place the inner radius of the reflecting 
region between 6 and 12 Rg and the outer radius at about 200 Rg. As {\saxj} is a 
pulsar, the inner radius can be interpreted as the magnetospheric radius, which 
is predicted by accretion theories to lie exactly between the coronation radius,
which in the case of {\saxj} is at about 30 km, and the NS surface in order to 
allow the observation of X-ray coherent pulsations from the source
\citep{Ghosh.Lamb:1979}. The statistics of the 2008 observation was nevertheless 
too low to discriminate between a symmetric and an asymmetric profile, although 
a disc interpretation is strongly favoured, as Compton broadening is not a viable 
explanation in a source whose Comptonized component originates at a large 
temperature \citep[$kT_e \ge 30$ keV, see e.g.][]{Gierlinski.etal:2002}. 

The main goal of this paper is to characterize the broad--band X-ray spectrum 
of the transient AMSP {\saxj}, and in particular the iron line and other reflection 
features, with a larger statistics than in previous observations, taking advantage 
from the large exposure of the 80~ks--{\xmm} observation and the broad--band
coverage provided by the {\nus} observation performed during the latest 
outburst from the source. This allowes us to acquire the source broadband spectrum 
and to constrain the reflection component properties such as the broad Fe emission 
line together with the expected Compton hump, therefore allowing to infer the 
properties of the accretion flow close to the NS. We report here on a detailed 
study of the reflection features and the fit, with a self-consistent 
reflection model, of both the iron line profile and the associated Compton 
reflection hump at energies above 10 keV. In this spectrum, which includes 
hard--band data (up to $50-70$ keV), the overall fractional amount of reflection is 
well determined by fitting the Compton hump. We can therefore test whether the 
observed iron line is consistent with this fractional amount of reflection. 
In this way we can confirm independently (fitting a different outburst state and
using different instruments) the inner disk parameters already obtained with 
{\xmm} and Suzaku for the 2008 outburst. 

\section{Observations}

\saxj, went into X-ray outburst in April 2015, after more than three years from 
its previous outburst in 2008. 
{\saxj} was observed by {\xmm} on 2015 April 11 (ObsID: 0724490201) for a total 
observing time of about 110 ks, as a result of an anticipated target of opportunity 
(ToO) observation approved to observe the source during an outburst. During the 
observation an abrupt drop-off of the count rate was visible in the EPIC/pn 
light curve caused by a problem with the Star Tracker, which led the satellite 
to be off-target for about seven hours, resulting in 80 ks effective on-source 
exposure. During the observation, the EPIC/pn camera was operated in timing mode 
to prevent photon pile-up and to allow the analysis of the coherent and 
aperiodic timing behaviour of the source \citep[see][for the timing analysis 
of these data]{Sanna.etal:2017b}. 
The EPIC/MOS cameras were switched off during the observation in order to 
allocate as much telemetry as possible to the pn in the case of high count 
rate, and the Reflection Grating Spectrometer (RGS) was operated in the standard 
spectroscopy mode. 


We have extracted source, background spectra and response matrices using the 
Science Analysis Software (SAS) v.16.1.0, setting the parameters of the tools accordingly. 
We produced a calibrated photon event file using reprocessing tools {\it epproc} and 
{\it rgsproc} for the pn and RGS data, respectively. Before extracting the spectra, 
we searched for contaminations due to background solar flares detected in the 
10-12 keV Epic-pn light-curve, but we did not find periods with high background. 

We also looked for the presence of pile-up in the pn spectrum; we have run the task 
{\it epatplot} and we did not find any significant contamination. The count-rate 
registered in the pn observation was around 450 c/s, which is below the limit 
for avoiding contamination by pile-up. Therefore, the source spectra were 
extracted from a rectangular region between $RAWX \geq 23$ and $RAWX \leq 49$. 
We selected only events with PATTERN $\leq 4$ and FLAG $= 0$ as a standard procedure 
to eliminate spurious events. We extracted the background spectrum from a region 
included between $RAWX \geq 5$ and $RAWX \leq 10$.  
Finally, using the task {\it rgscombine} we have obtained the added source 
spectrum for RGS1+RGS2, the relative added background spectrum along with the 
relative response matrices. 
We have fitted RGS spectrum in the 0.5-1.8 keV energy range, whereas the pn 
spectrum in the 2.4-10 keV energy range. 
The spectral analysis of the {\xmm}/EPIC-pn spectrum was restricted to 2.4-10 keV to 
exclude the region around the detector Si K-edge (1.8 keV) and the mirror Au M-edge 
(2.3 keV) that could affect our analysis, as well as to exclude possible residuals 
of instrumental origin below 2 keV that usually appear in case of bright sources 
observed in timing mode \citep[see e.g.][]{D_Ai.etal:10, Egron.etal:13}. 

In this paper we also analyze data collected by the {\nus} satellite. 
A ToO was requested to observe the source during the 2015 outburst in order 
to complement the {\xmm} spectrum with high energy coverage. 
The {\nus} observation, obtained as Director Discretionary Time (DDT), 
was performed four days after the {\xmm} observation, on 2015 April 15
(ObsID: 90102003002), for a total observing time of 55 ks, resulting 
in roughly 49 ks of exposure per telescope. Science data were extracted 
using NuSTARDAS (NuSTAR Data Analysis Software) v1.7.1. 
Source data have been extracted from a circular region with 120"  
radius whereas the background has been extracted from a circular 
region with 60" radius in a position far from the source. 
With the aim to get "STAGE 2" events clean, we run the {\it nupipeline} 
with default values of the parameters and with the parameter {\it SAAMODE} 
set to {\it optimized} in order to eliminate high background events caused 
by the SAA passage. The average count rate during the {\nus} observation 
was $\sim 35-40$ c/s. 

A type-I burst is present during the {\nus} observation,
at about 14 ks from the beginning of the observation. The burst profile is
not complete since the rise phase was in coincidence with a gap in the 
light curve. The peak of the burst seems to reach approximately 200 c/s,
about a factor 4 the level of the persistent emission, and it lasted about
200 s. We eliminated a time interval of 250 s starting from 5 s before the 
rise of the burst, and checked that the spectra did not change significantly.
Spectra for both 
detectors, FPMA and FPMB, were extracted using the {\it nuproducts} 
command. Corresponding response files were also created as output 
of {\it nuproducts}. A comparison of the FPMA and FPMB spectra, indicated 
a good agreement between them. We have therefore created a single 
added spectrum, with its corresponding background spectrum, ancillary response
file and matrix response, using the {\it addascaspec} command. 
In this way, we obtain a summed spectrum for the two {\nus} modules 
\citep[see e.g.][]{Miller.etal:13}. We fitted this spectrum in the 3-70 keV 
energy range, where the emission from the source dominates over the background. 

\begin{figure}
\begin{center}
    \begin{minipage}{16cm}
      \includegraphics[width=10cm]{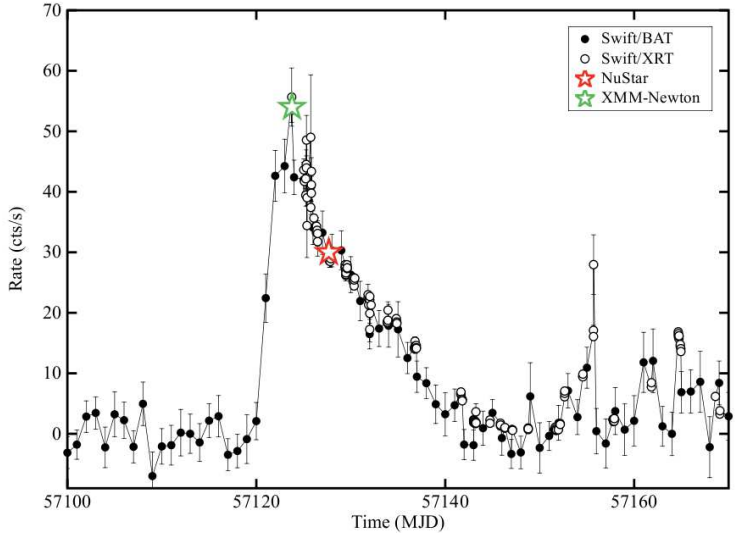}
    \caption{Swift/BAT and XRT light curve during the 2015 outburst of {\saxj}.
             The dates of the {\xmm} and {\nus} observations are indicated with stars.}
    \label{lcurve}
    \end{minipage}
\end{center}
\end{figure}

In Figure \ref{lcurve} we show the light curve during the 2015 outburst of {\saxj}
obtained with the instruments, XRT and BAT, on board the Swift satellite. 
In this light curve the dates of the {\xmm} and {\nus} observations are indicated 
with stars.

\section{Spectral Analysis and Results}

For spectral analysis, the EPIC/pn energy channels were grouped in order to have
at least 20 counts per energy channel and to oversample the energy resolution 
element by no more than three channels. RGS and {\nus} spectra were grouped in 
order to have at least 20 counts per energy channel. 
The X-ray spectral package we use to model the observed emission is 
XSPEC v.12.9.1.  
For each fit we have used {\it phabs} in XSPEC to model the photoelectric 
absorption due to neutral matter, with photoelectric cross sections from 
\citet{Balucinska:1992} and element abundances from \citet{Anders:1989}.

\subsection{The {\nus} spectrum}
\begin{figure}
\begin{center}
    \begin{minipage}{16cm}
      \includegraphics[width=6cm,angle=270]{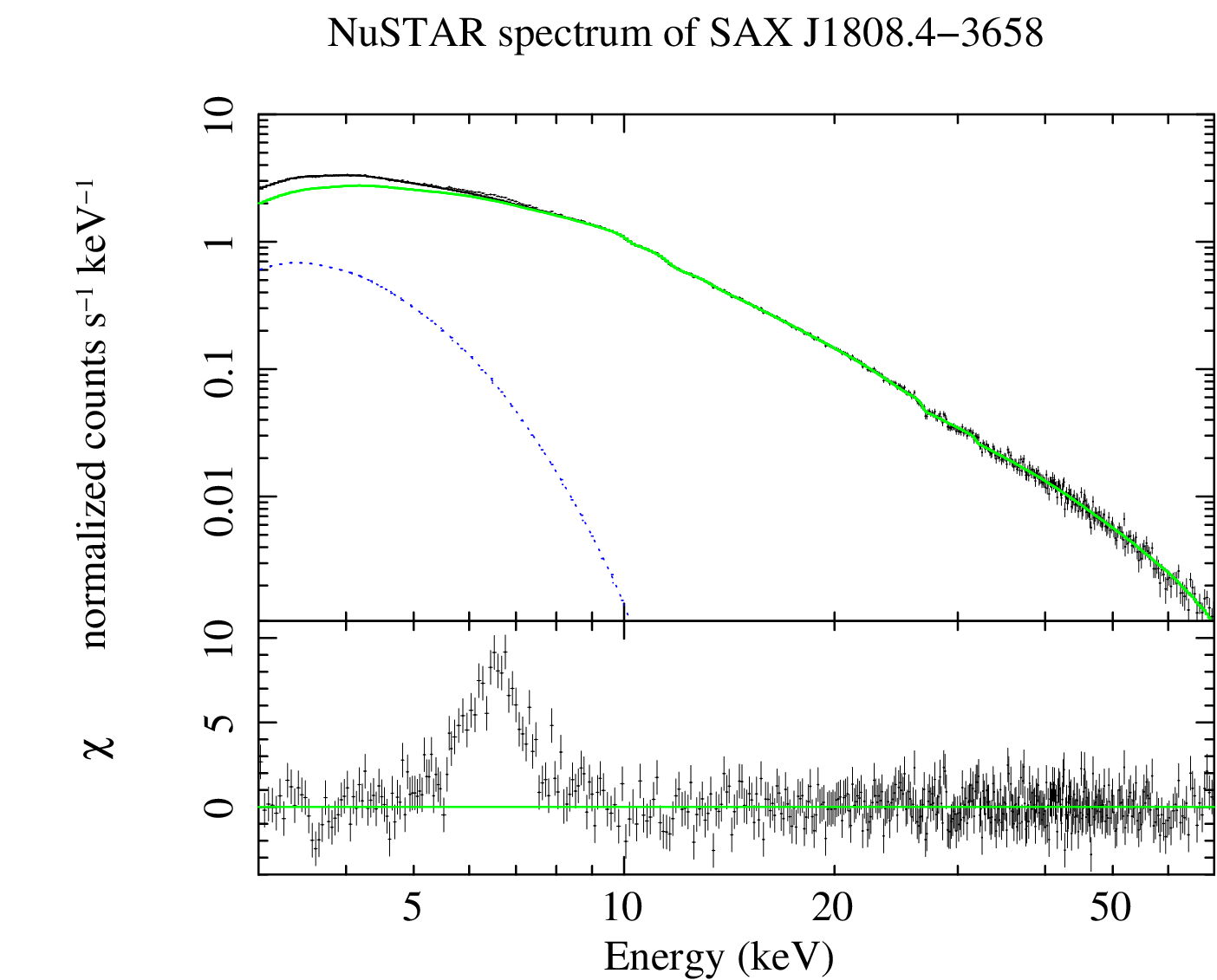}
      \includegraphics[width=6cm,angle=270]{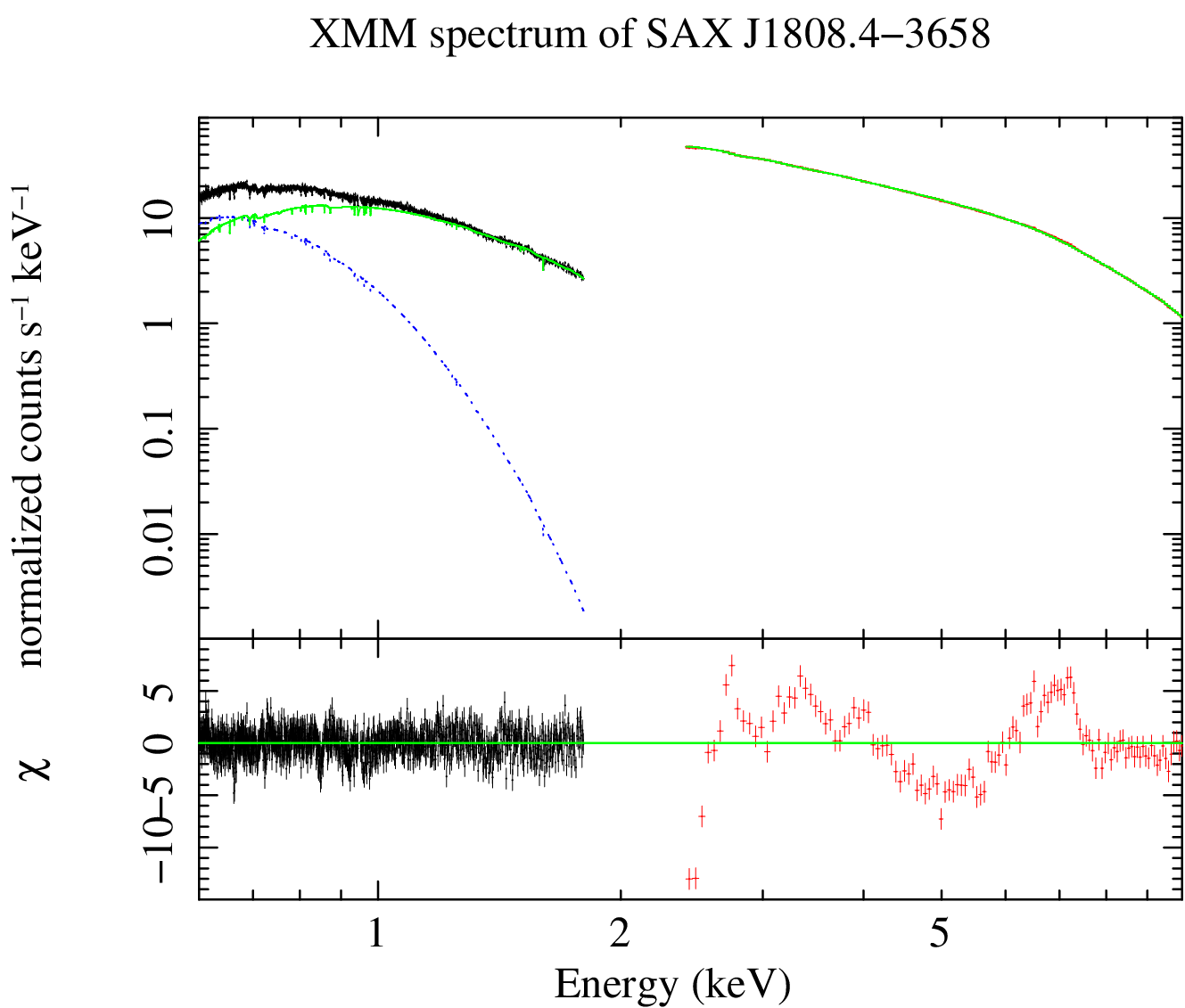}
    \caption{{\bf Left panel:}{\nus} spectrum in the energy range 3 - 70 keV (top) 
    and residuals in units of $\sigma$ (bottom) with respect to the continuum model  
    of {\saxj} when the Fe line normalization is set to 0. 
    {\bf Right panel:} {\xmm} spectrum in the energy range 0.6 - 10 keV (top) 
    and residuals in units of $\sigma$ (bottom) with respect to the continuum model  
    of {\saxj}. In both cases the model consists of a blackbody (dotted line) and the 
    Comptonization component {\it nthComp} (solid line), both multiplied by the photoelectric 
    absorption component modeled with {\it phabs}. Evident residuals are present at the 
    expected iron line energy of $6.4-6.7$ keV. Residuals between 2.4 and 4.5 keV are
    also present in the pn spectrum.}
    \label{nuspec}
    \end{minipage}
\end{center}
\end{figure}

We have started analyzing the broad-band spectrum acquired with {\nus} 
in the energy range $3-70$ keV. We fit the continuum emission with a 
soft blackbody and the Comptonization model {\it nthComp} in XSPEC
\citep{Zycki.etal:99}, modified at low energy by photoelectric absorption 
caused by neutral matter. 
This is a standard model to fit the broad-band continuum emission in NS 
LMXBs of the atoll class both in the soft and in the hard states 
\citep[see e.g.][and references therein]{Piraino.etal:07, Di_Salvo.etal:09, 
Egron.etal:13, Sanna.etal:2013, DiSalvo.etal:2015} and provides also a good 
fit to the broad-band continuum emission in AMSPs \citep[see e.g.][]{Papitto.etal:10, 
Papitto.etal:13b, Papitto.etal:2016, Sanna.etal:2016}. 
Because of the lack of sensitivity of {\nus} at low energies, we had to fix
the photoelectric equivalent hydrogen column density at $0.21 \times 10^{22}$ 
cm$^{-2}$, which is the best fit value for this parameter obtained 
for the X-ray spectrum of {\saxj} observed by {\xmm} during the 2008
outburst (see P09). The soft blackbody component has a best fit temperature 
of $\sim 0.7$ keV, the seed photons for Comptonization, assumed to have a 
blackbody spectrum, have a temperature lower than $\sim 0.3$ keV; these photons 
are comptonized in a hot corona with an electron temperature of $\sim 30$ keV 
and a moderate optical depth, corresponding to a photon index of the Comptonization 
spectrum of $\Gamma \sim 1.8$.
This continuum model gave, however, an unacceptable fit, corresponding to 
a $\chi^2 / dof = 1679 / 1325$, because of the presence of evident 
localized residuals in the $5-9$ keV range, clearly caused by the presence
of a broad iron line, whose profile is evident in Figure~\ref{nuspec}
(left panel). 

To fit these residuals we first added to our continuum model a {\it Gaussian}
component to model the broad iron line profile visible in the residuals. The addition 
of this component resulted in a significant improvement of the fit and to an 
acceptable $\chi^2 / dof = 1316 / 1322$. The Gaussian centroid was at $6.51 \pm
0.08$ keV, its width $\sigma = 0.65 \pm 0.10$ keV, and its equivalent width was
$101 \pm 13$ eV. We also tried to fit the residuals at the iron line energy with 
a {\it diskline} profile \citep{Fabian.etal:89} instead of a Gaussian. 
The addition of this component resulted in a further improvement of the fit, 
$\chi^2 / dof = 1306 / 1319$. 
The {\it diskline} component is characterized by a centroid energy of 
$6.3-6.5$ keV, indicating neutral and/or weakly ionized iron (Fe I-XV),
an emissivity index, describing the emissivity of the disk as a function 
of the emission radius $\propto r^{Betor}$, of $Betor \sim -(1.8 - 2.2)$,
an inner and outer radius of the emitting region of $R_{in} \le 7$ R$_g$
and $R_{out} \sim 290 - 1100$ R$_g$, respectively, where R$_g = G M_{NS}/c^2$ 
is the gravitational radius, and a large inclination angle of the system
with respect to the line of sight, $i \ge 70^\circ$. The best fit parameters 
are reported in Table \ref{Tab1}. 
These parameters are very similar (compatible well within the errors) to the 
best fit parameters of the iron line component obtained from the {\xmm} 
spectrum of {\saxj} during the 2008 outburst (cf.\ P09); in particular the 
previous {\xmm} spectrum also gave a quite small outer disk radius of 
$140 - 360$ R$_g$ and a high inclination angle of $i \ge 60^\circ$.

The broad iron line profile seems to be compatible with reflection of
the main Comptonization spectrum off the accretion disk, where the broadness
of the profile is induced by the fast motion of the matter in the inner
disk, and related (mildly) relativistic effects. However, if the iron line 
is produced by reflection then a Compton hump should be visible in the 
high energy part of the spectrum given that the source is in a hard 
state. To test this hypothesis we tried to substitute the 
{\it diskline} component with a self-consistent reflection model to fit 
both the iron line and the Compton hump.
In particular, we used the {\it relxillCp} model \citep{Garcia.etal:2014}, 
which models the irradiation of the accretion using an {\it nthComp} 
Comptonization continuum (Cp). Note that at the moment it is not possible to 
fit the temperature of the seed photons for the Comptonization component, 
that is fixed at 50 eV (Thomas Dauser, private communication). This should be
anyway a good approximation for the case of {\saxj} given the low value we find 
for this temperature (see Table \ref{Tab1} and \ref{Tabtot}). 

The {\it relxillCp} model
allows to determine the reflection fraction, defined as ratio of intrinsic 
intensity emitted towards the disk compared to that escaping to infinity 
\citep[see][for more details]{Dauser.etal:2016}, the
inclination angle of the system and the ionization parameter of the disk,
given by $\log \xi$, where $\xi = L_X /(n_e r^2)$, where $L_X$ is the 
luminosity of the incident X-ray spectrum, $n_e$ is the electron number
density in the emitting region, and $r$ is the distance between the 
illuminating source and the emitting region. This reflection model
includes the smearing component {\it relconv}\footnote{More details 
can be found in the following webpage: 
http://www.sternwarte.uni-erlangen.de/~dauser/research/relxill/} in order 
to take into account Doppler and relativistic effects caused by the fast 
motion of the matter in the disk. 
Unless specified otherwise, the two emissivity indices (defined in order to be
negative as in the case of the {\it diskline} parameter {\it Betor}) have been 
constrained in order to assume the same value and the dimension-less spin 
parameter a has been fixed to 0.

This model gives a good fit to the data, corresponding to a $\chi^2 / 
dof = 1300/1320$. We find that fixing the iron abundance, $A_{Fe}$, at 2 
times the solar value the fit slightly improved returning a  
$\chi^2 / dof = 1297 / 1320$, lower with respect to the previous 
fit with a {\it diskline}, corresponding to a $\Delta \chi^2 = 9$
with one extra degree of freedom. Letting the iron abundance 
free to vary, we find that its best fit value was $\sim 3$ corresponding
to a $\chi^2$ very similar to the previous one, but the uncertainty on 
this parameter is quite large ($A_{Fe} \simeq 1 - 4$), so we preferred 
to keep fixed this parameter at 2 times the solar value. The outer disk radius
has been left free to vary, but its uncertainty could not be determined because
the $\chi^2$ was quite insensitive to its value; probably the energy
resolution of {\nus} is not enough to constrain this parameter.
The spectral results are reported in Table \ref{Tab1}, and in Figure \ref{fig3} 
(right panel) we show the {\nus} spectrum, the best-fit model, and the residuals 
in units of $\sigma$ with respect to the best-fit model. We extrapolated the 
total $0.5-200$ keV observed luminosity of the source during the {\nus} 
observation, corresponding to $(3.78 \pm 0.15) \times 10^{36}$ ergs/s assuming 
a distance to the source of 3.5 kpc \citep{Galloway.etal:2006}.

\begin{table}
\caption{The best fit parameters of the spectral fitting of the {\nus} ($3-70$
keV energy band) spectrum of {\saxj}. 
In all the cases the continuum emission is described by a combination of 
a blackbody and the Comptonization component {\it nthComp}, modified at
lower energy by photoelectric absorption from neutral matter modeled with
{\it phabs}. The reflection component is fitted with a {\it diskline} component,
or with the self-consistent reflection model {\it relxillCp}.
The blackbody luminosity is given in units of $L_{36}/D_{10}^2$,
where $L_{36}$ is the bolometric luminosity in units of $10^{36}$~ergs/s and 
$D_{10}$ the distance to the source in units of 10~kpc. 
The blackbody radius is calculated in the hypothesis of spherical emission 
and for a distance of 3.5~kpc. 
Smearing indicate the smearing component of the {\it diskline} and 
{\it relxillCp} models, respectively. 
Flux in the nthComp component is calculated in the $1-10$ keV range, while 
total flux is calculated in the $1.6-70$ keV band. 
Uncertainties are given at $90\%$ confidence level. INDEF means that the error
on the parameter could not be calculated being the $\chi^2$ quite insensitive to
its value.}
\begin{tabular}{|llcc|}
\hline
             &           & DISKLINE & RELXILLCP     \\
Component    & Parameter & {\nus} &  {\nus}  \\
             &           & ($3-70$ keV) &  ($3-70$ keV)  \\
\hline
phabs & $N_H$ ($\times 10^{22}$ cm$^{-2}$) & $0.21$ (fixed) & $0.21$ (fixed)   \\
bbody & $kT_{BB}$ (keV)          & $0.697 \pm 0.015$ &  $0.672 \pm 0.007$   \\
bbody & L$_{BB}$ ($L_{36}/D_{10}^2$)  & $2.14^{+0.09}_{-0.04}$   &  $1.79 \pm 0.04$  \\
bbody & R$_{BB}$ (km)                 &  $2.93 \pm 0.14$  &  $2.88 \pm 0.07$  \\
nthComp & $kT_{seed}$ (keV)		& $< 0.29$  & $-$    \\
nthComp& $\Gamma$                     & $1.819 \pm 0.006$ &  $1.868 \pm 0.015$  \\
nthComp & $kT_e$ (keV)			& $26.1^{+2.3}_{-1.8}$ & $39^{+13}_{-3}$   \\
nthComp & Flux ($10^{-10}$ ergs cm$^{-2}$ s$^{-1}$) &    $9.5 \pm 1.4$  &    $-$  \\
diskline & $E_{line}$ (keV)	        & $6.38 \pm 0.10$  & $-$     \\
diskline & $I_{line}$ ($10^{-4}$ ph cm$^{-2}$ s$^{-1}$)  & $8.6 \pm 1.2$  &  $-$    \\
diskline & $EqW$ (eV)	               & $124 \pm 19$  &  $-$     \\
Smearing & $Betor$     & $-2.04^{+0.19}_{-0.15}$  & $-1.95 \pm 0.12$     \\  
Smearing & $R_{in}$ ($G M / c^2$)                 & $< 7$ & $14.9 \pm 2.5$    \\  
Smearing & $R_{out}$ ($G M / c^2$)  & $520^{+550}_{-240}$ & $1000$ (INDEF)  \\ 
Smearing & Incl (deg)               & $> 70$ & $50^{+22}_{-5}$ \\  
RelxillCp & Refl Frac               & $-$  &  $0.62 \pm 0.04$   \\  
RelxillCp & Fe abund                & $-$  &  $2$ (fixed)    \\  
RelxillCp & $\log \xi$              & $-$  &  $2.76^{+0.10}_{-0.07}$    \\  
RelxillCp & Norm ($\times 10^{-3}$)	& $-$ & $3.73^{+0.07}_{-0.14}$   \\
\hline
total & Flux ($10^{-9}$ ergs cm$^{-2}$ s$^{-1}$)  & $2.01 \pm 0.12$  & $2.02 \pm 0.09$   \\
total &  $\chi^2$ (dof)	   & $1305.68~(1319)$ &  $1296.66~(1320)$    \\
\hline \\
\end{tabular}
\label{Tab1}
\end{table}

\subsection{The {\xmm} spectrum}

In order to check the results obtained from the {\nus} spectrum, we have 
fitted separately the {\xmm} spectra (RGS and pn, energy range $0.6-10$ keV), 
using the same continuum model. 
Again, fitting only the continuum results in an unacceptable fit,
with $\chi^2/dof = 3489/1458$, and clear residuals are present at the 
iron line energy and lower energies, at $\sim 2.6$ keV, $\sim 3.3$ keV,
and $\sim 4$ keV (see Fig. \ref{nuspec}, right panel). These low-energy
residuals are similar to those observed in other bright LMXBs of the
atoll class in the soft state, as 4U 1705-44 \citep[see e.g.][]{Di_Salvo.etal:09, 
Egron.etal:13} or GX 3+1 \citep[e.g.][]{Piraino.etal:12, Pintore.etal:2015}. 
We therefore added to the continuum model four {\it diskline} components to 
fit the iron line and the other low energy features. The smearing parameters 
of the {\it diskline} fitting the low-energy lines were fixed to be the same 
of those of the {\it diskline} fitting the iron line. In this way we get a 
significative improvement of the fit, corresponding to $\chi^2/dof = 2094/1446$.
The best fit parameters of this fit are reported in Table \ref{Tabtot}, and
in Figure \ref{fig3} (left panel) we show the {\xmm} spectrum, the best-fit 
model, and the residuals in units of $\sigma$ with respect to the best-fit model.

Note that the electron temperature of the Comptonization component is much
lower with respect to that measured with {\nus} four days later, indicating
that {\saxj} could have been in a soft state at the time of the {\xmm} 
observation, in agreement with the presence of ionized discrete features 
in the spectrum. Note also that the {\xmm} observation was acquired at the
peak of the outburst and this is the first time that the {\saxj} spectrum 
has been observed with good energy resolution 
at the peak of an outburst. 
Substituting the {\it diskline} used for the iron line with the reflection
model {\it relxillCp},  
keeping the smearing parameters fixed to the corresponding smearing 
parameters of the other, low-energy {\it disklines}, returns a 
$\chi^2/dof = 2100/1447$, that is slightly worse 
than before ($\Delta \chi \simeq 6$), but with one extra degree of freedom. 
All the best-fit parameters are consistent within the errors with the 
best-fit values of the previous fit with disklines.
As regards the reflection parameters, we fixed the iron abundance to 2, as
in the case of the {\nus} spectrum, the reflection fraction is compatible 
to that obtained with {\nus}, although with a larger uncertainty, while
the ionization parameter results to be much higher, $\log \xi = 3.6 - 3.8$,
in agreement with the presence of features from highly ionized elements in
the {\xmm} spectrum. 
The best fit parameters of this fit are reported in Table \ref{Tabtot}.
Note that some residuals are still present at $\sim 7$ keV, this is also
visible in Figure \ref{fig3}. To fit these residuals we tried to add a 
Gaussian emission line at that energy, obtaining a $\chi^2/dof = 2072/1444$;
the improvement of the fit is barely significant, corresponding to an F-test
probability of chance improvement of $\sim 1.2 \times 10^{-4}$. We also 
tried to fit the iron abundance obtaining a preference for an overabundance,
$A_{Fe} = 3.4 \pm 0.7$, but without a statistically significant improvement
of the fit (F-test probability of chance improvement $\sim 1.3 \times 10^{-3}$
for the addition of one parameter).

\subsection{Combined analysis of the {\nus} and {\xmm} spectra}

In order to increase the statistics at the iron line energy and for the 
whole reflection component, we tried to fit together
the {\nus} and {\xmm} spectra, using the best-fit models obtained above. 
The two observations are not perfectly simultaneous, the {\nus}
observation being performed four days after the {\xmm} observation 
taken at the peak of the outburst. Both the best fit continuum emission
and the emission lines are very different between the two spectra.
We therefore left most of the parameters free to vary between the
two spectra and only few parameters were constrained to assume the
same value for the two spectra; these are the equivalent hydrogen 
column density, $N_H$ of the interstellar absorption, all the parameters 
of the smearing component and the iron abundance.  
Both the ionization parameter of the reflection component and the reflection
fraction were quite different between the {\xmm} and {\nus} spectra (cf.\
Table \ref{Tab1} and Table \ref{Tabtot}) and therefore we let these parameters
free to vary between the two spectra.
The low-energy disklines are not required by the {\nus} spectrum and
hence are not included in the fit of the {\nus} spectrum.
Again, letting the iron abundance free to vary (but forced to assume the same 
value for the two spectra) we find an improvement of the fit, corresponding to 
a $\chi^2/dof = 3394.30/2770$ (corresponding to an F-test probability of chance 
improvement of $4 \times 10^{-5}$ for the addition of one parameter). The best-fit 
value was $A_{Fe} \simeq 3.4$, while all the other parameters did not change 
significantly. However, we could not determine the error on this parameter, 
and therefore we preferred to keep the iron abundance fixed to two times the 
Solar value.
In Table \ref{Tabtot} (last column) we report the best-fit parameters obtained 
for this fit. In Figure \ref{fig4} we show the {\xmm} and {\nus} spectra together 
with the best fit model and residuals in units of sigma with respect to
this model. 

\begin{table}
\caption{The best fit parameters of the spectral fitting of the {\xmm} ($0.6-10$
keV energy band) and {\xmm} + {\nus} ($0.6-70$ keV energy band) spectra of {\saxj}. 
In all the cases the continuum emission is described by a combination of 
a blackbody and the Comptonization component {\it nthComp}, modified at
lower energy by photoelectric absorption from neutral matter modeled with
{\it phabs}. The reflection component is fitted with {\it diskline} components
or with the self-consistent reflection model {\it relxillCp}.
The blackbody luminosity is given in units of $L_{36}/D_{10}^2$,
where $L_{36}$ is the bolometric luminosity in units of $10^{36}$~ergs/s and 
$D_{10}$ the distance to the source in units of 10~kpc. The blackbody radius
is calculated in the hypothesis of spherical emission and for a distance of
3.5~kpc. 
Fluxes in the nthComp component are calculated in the $1-10$ keV range, while 
total flux is calculated in the $0.6-10$ keV band for the {\xmm} spectrum and 
in the $1.6-70$ keV band for the {\nus} spectrum. 
Uncertainties are given at $90\%$ confidence level.}
\begin{tabular}{|llcccc|}
\hline
             &           & DISKLINE & RELXILLCP & \multicolumn{2}{|c|}{RELXILLCP}   \\
  Component    & Parameter & {\xmm} & {\xmm} &  {\xmm} & {\nus} \\
             &           & ($0.6-10$ keV) & ($0.6-10$ keV) & \multicolumn{2}{|c|}{($0.6-70$ keV)} \\
\hline
phabs & $N_H$ ($\times 10^{22}$ cm$^{-2}$) & $0.164 \pm 0.006$ & $0.142 \pm 0.011$  &  \multicolumn{2}{|c|}{$0.146 \pm 0.011$}  \\
bbody & $kT_{BB}$ (keV)  & $0.103 \pm 0.002$ &  $0.109 \pm 0.003$  & $0.112^{+0.003}_{-0.002}$ & $0.680^{+0.002}_{-0.003}$ \\
bbody & L$_{BB}$ ($L_{36}/D_{10}^2$)  & $10.3^{+0.9}_{-0.6}$   &  $4.4^{+1.2}_{-0.9}$  & $5.4 \pm 1.1$ & $1.64 \pm 0.11$ \\
bbody & R$_{BB}$ (km)                 &  $295 \pm 17$  &  $172 \pm 25$ &  $180 \pm 20$ & $2.69 \pm 0.09$ \\  
nthComp & $kT_{seed}$ (keV)	      & $0.15^{+0.02}_{-0.04}$  &  $-$  & \multicolumn{2}{|c|}{$-$}  \\
nthComp& $\Gamma$                     & $1.904 \pm 0.008$ & $1.78 \pm 0.04$ & $1.73 \pm 0.03$ & $1.88 \pm 0.02$  \\
nthComp & $kT_e$ (keV)		      & $4.7 \pm 0.3$ &  $7.2 \pm 0.8$  & $6.2 \pm 0.5$ & $36.9^{+6.4}_{-4.6}$   \\
nthComp & Flux ($10^{-10}$ ergs cm$^{-2}$ s$^{-1}$) &    $14.7 \pm 0.9$  & $-$ & $-$  &  $-$  \\   
diskline & $E_{line}$ (keV)	      & $2.665 \pm 0.012$  & $2.666 \pm 0.015$ & $2.664 \pm 0.016$ & $-$  \\
diskline & $I_{line}$ ($\times 10^{-4}$ ph cm$^{-2}$ s$^{-1}$)  & $21.5 \pm 1.9$  & $21 \pm 3$ & $18 \pm 3$ &  $-$ \\
diskline & $EqW$ (eV)	               & $41 \pm 6$  &  $42 \pm 6$ &  $35 \pm 11$  & $-$  \\   
diskline & $E_{line}$ (keV)	        & $3.294 \pm 0.019$  & $3.31 \pm 0.02$ & $3.28 \pm 0.02$ & $-$ \\
diskline & $I_{line}$ ($\times 10^{-4}$ ph cm$^{-2}$ s$^{-1}$)  & $15.0 \pm 1.5$  &  $16 \pm 3$  &  $11.6 \pm 1.6$  & $-$ \\
diskline & $EqW$ (eV)	               & $38 \pm 6$  &  $50 \pm 8$  &  $33 \pm 11$  & $-$ \\    
diskline & $E_{line}$ (keV)	        & $3.99 \pm 0.04$  & $4.06 \pm 0.04$ & $3.99 \pm 0.05$  & $-$ \\
diskline & $I_{line}$ ($\times 10^{-4}$ ph cm$^{-2}$ s$^{-1}$)  & $8.0^{+0.7}_{-0.5}$  &  $9.5 \pm 1.3$  & $6.3^{+1.1}_{-0.9}$  &  $-$ \\
diskline & $EqW$ (eV)	               & $30 \pm 4$  &  $42 \pm 7$  &  $31 \pm 9$  &  $-$ \\   
diskline & $E_{line}$ (keV)	        & $6.70 \pm 0.05$  & $-$ & $-$  &  $-$  \\
diskline & $I_{line}$ ($\times 10^{-4}$ ph cm$^{-2}$ s$^{-1}$)  & $5.4 \pm 0.6$  &  $-$  &  $-$  &  $-$  \\
diskline & $EqW$ (eV)	               & $58 \pm 8$  &  $-$  &  $-$  &  $-$   \\   
Smearing & $Betor$     & $-2.16 \pm 0.10$  &  $-2.20 \pm 0.04$  &  \multicolumn{2}{|c|}{$-2.08^{+0.03}_{-0.06}$}    \\  
Smearing & $R_{in}$ ($G M / c^2$)       & $10^{+8}_{-3} $ &  $<7.2$  &  \multicolumn{2}{|c|}{$9.9^{+1.6}_{-2.4}$}   \\  
Smearing & $R_{out}$ ($G M / c^2$)  & $1032^{+750}_{-270}$ & $>900$ & \multicolumn{2}{|c|}{$> 850$}   \\ 
Smearing & Incl (deg)               & $> 58$  & $59.8^{+4.2}_{-1.6}$ & \multicolumn{2}{|c|}{$67^{+9}_{-8}$} \\  
RelxillCp & Refl Frac               & $-$  &  $0.69^{+0.27}_{-0.19}$  &  $0.9^{+0.4}_{-0.3}$  &  $0.22^{+0.11}_{-0.04}$  \\  
RelxillCp & Fe abund                & $-$  &  $2$ (fixed) &  \multicolumn{2}{|c|}{$2$ (fixed)}   \\  
RelxillCp & $\log \xi$              & $-$  &  $3.72 \pm 0.09$  &  $3.78 \pm 0.08$  &  $2.4 \pm 0.3$  \\  
RelxillCp & Norm ($\times 10^{-3}$) & $-$  &  $2.64^{+0.25}_{-0.36}$  &  $2.4 \pm 0.2$   &   $3.79^{+0.10}_{-0.07}$  \\  
\hline
total & Flux ($10^{-9}$ ergs cm$^{-2}$ s$^{-1}$) & $1.59 \pm 0.02$  &  $1.60 \pm 0.07$  &  $1.59 \pm 0.22$  &  $2.03 \pm 0.04$   \\   
total &  $\chi^2$ (dof)	   & $2094.07~(1446)$ & $2100.49~(1447)$ & \multicolumn{2}{|c|}{$3415.08~(2771)$}   \\
\hline \\
\end{tabular}
\label{Tabtot}
\end{table}


\begin{figure}
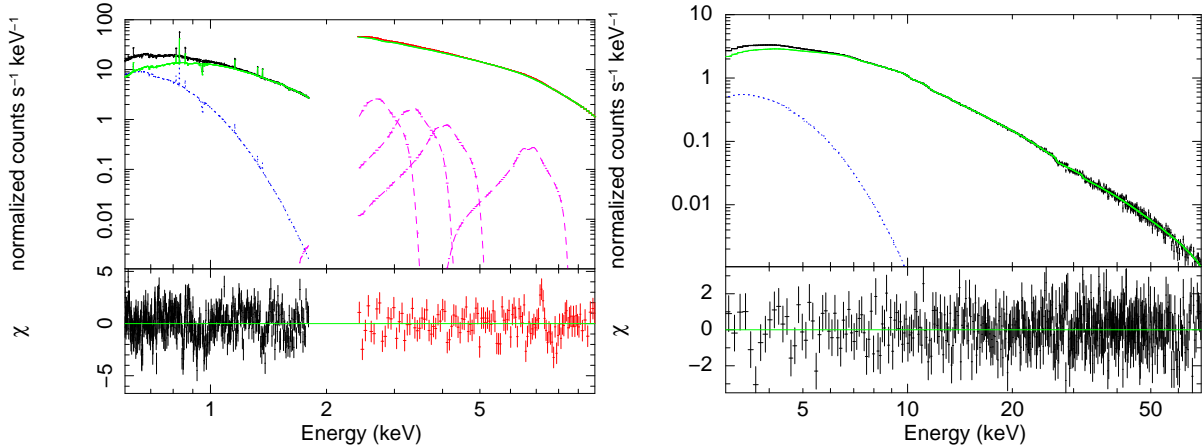

\begin{center}
    \begin{minipage}{16cm}
      \includegraphics[width=6cm,angle=270]{XMM_res_disklines.ps}
      \includegraphics[width=6cm,angle=270]{NuSTAR_data_res_relxill_Fe2.ps}
    \caption{{\bf Left:} {\xmm} RGS (black points) and pn (red points) spectra 
    of {\saxj} in the energy range 0.6 - 10 keV (top) and residuals in units of 
    $\sigma$ (bottom) with respect to the best-fit model (see Table~\ref{Tabtot}, 
    third column). 
    The model consists of a blackbody (dotted line), the Comptonization component 
    {\it nthComp} (solid line) and four disklines (dashed lines) describing the 
    reflection component, all multiplied by photoelectric absorption.
        {\bf Right:} {\nus} spectrum in the energy range 3 - 70 keV (top)
    and residuals in units of $\sigma$ (bottom) with respect to the best-fit model 
    shown in the last column of Table~\ref{Tab1}. The model
    components are also shown. From the left to the right we see the blackbody
    component (dotted line), the Comptonization component plus the 
    smeared reflection component modeled by {\it relxillCp} (solid line). }
    \label{fig3}
    \end{minipage}
\end{center}
\end{figure}

\begin{figure}
\begin{center}
    \begin{minipage}{16cm}
      \includegraphics[width=12cm,angle=270]{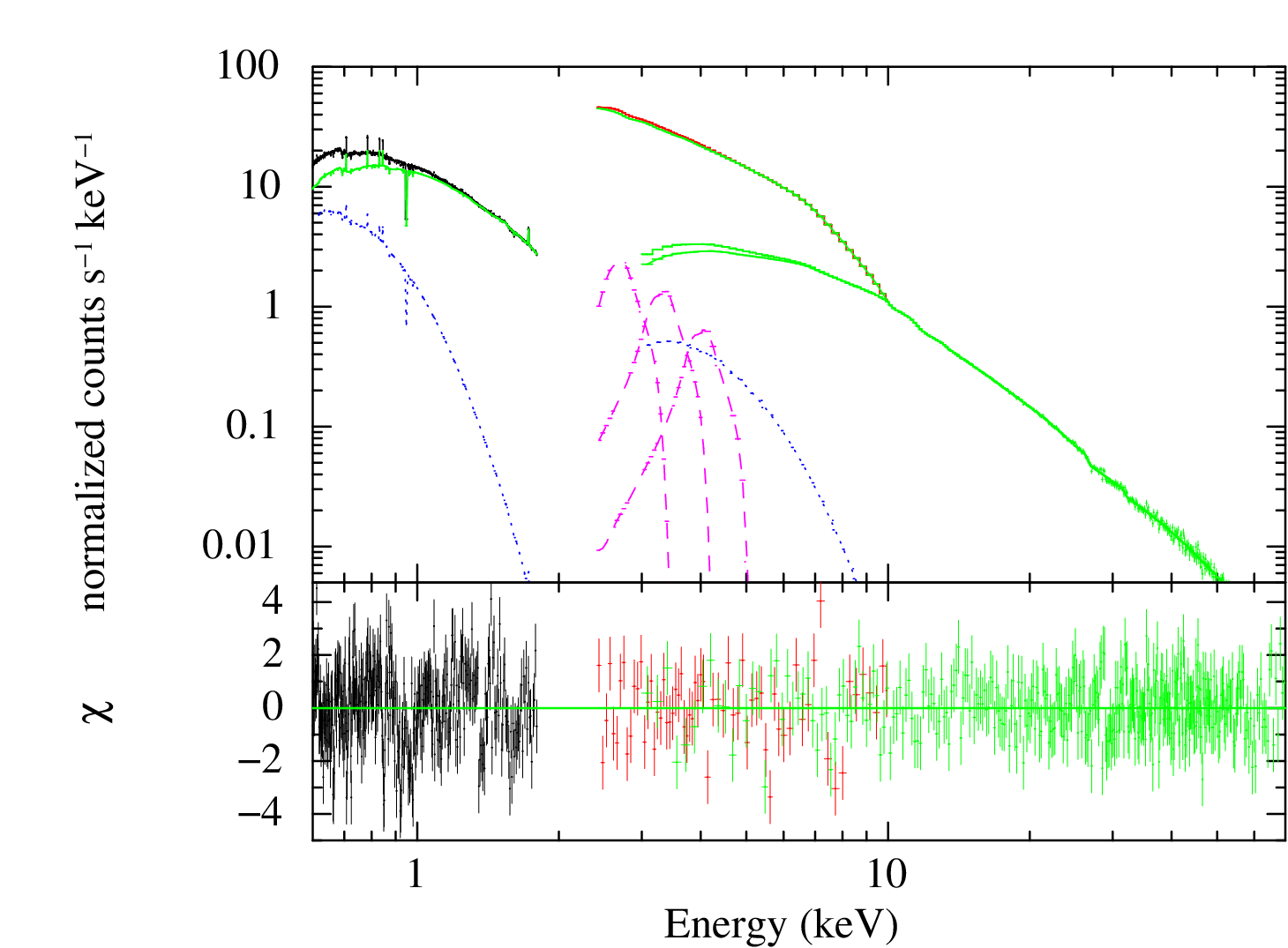}
      \includegraphics[width=4.8cm,angle=270]{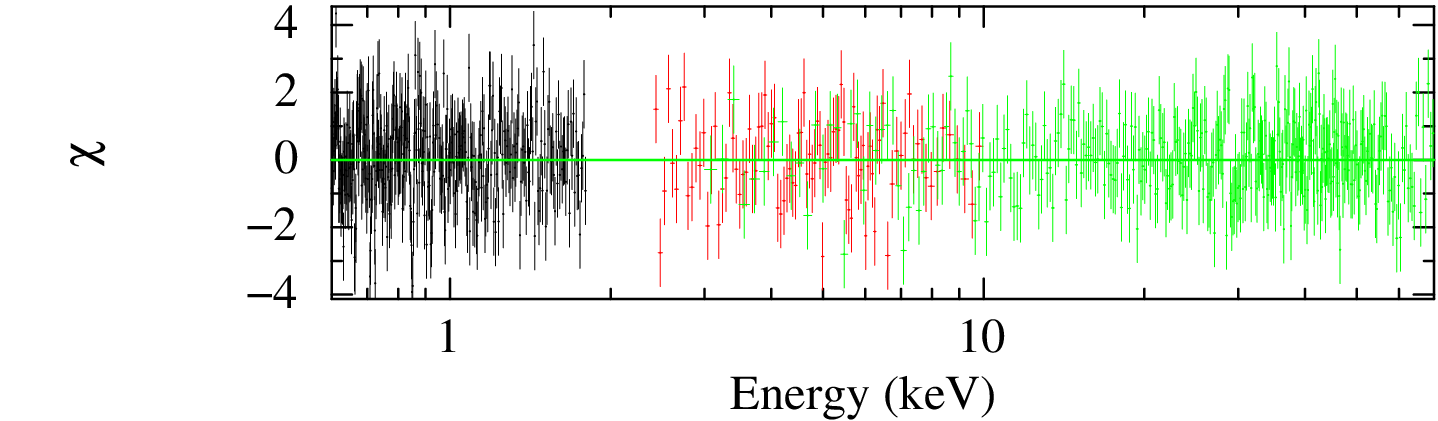}
    \caption{{\xmm} (black and red points) and {\nus} (green points) spectra of {\saxj} 
    (top) and residuals in units of $\sigma$ (middle) 
    with respect to the best-fit model (see Table~\ref{Tabtot}, last column). 
    The model consists of a blackbody (with different temperatures for the {\xmm} and 
    {\nus} spectra, dotted lines), the Comptonization component plus the 
    smeared reflection component modeled by {\it relxillCp} (solid line), all multiplied 
    by photoelectric absorption. Three disklines (indicated with 
    dashed lines) are used to fit the {\xmm} spectra but are not required for the 
    {\nus} spectrum. The total model 
    is plotted on top of the data. Note that each spectrum is convolved with its response
    matrix and effective area, as well as the corresponding model and model components.
    In the bottom panel we show the residuals in units of $\sigma$ with respect to the 
    best-fit model including the absorption components (see Tab.\ \ref{Tabfin}).}
    \label{fig4}
    \end{minipage}
\end{center}
\end{figure}

Form visual inspection of Fig.\ \ref{fig3} and \ref{fig4} some residuals are
still evident in the {\xmm} RGS (at $\sim 0.9$, $\sim 1.3$ and $\sim 1.6$ keV) and pn 
spectra (at $\sim 7$ keV). In order to fit these residuals we tentatively add
to the previous model three (Gaussian) absorption lines and an edge. The edge 
has a best-fit energy of $\sim 7.4$ keV and may be associated to mildly ionized 
iron (Fe V-XV). The centroid energies of the three absorption lines are 0.947 keV 
(possibly from Ne IX, resonance transition rest-frame energy 0.922 keV), 1.372 keV 
(possibly from Mg XI, resonance transition rest-frame energy 1.352 keV) and 1.596 
keV, respectively.
The latter is remarkably close to the resonant line of Al XII (rest-frame 1.598 keV).
Note, however, that this element has a low cosmic abundance ($\sim 3 \times 10^{-6}$ 
in number of atoms with respect to H) and that this line has a significance of about $3.5 \sigma$;
there is the possibility that this line is of instrumental origin, and we prefer not 
to discuss it further. 

The first two absorption lines appear, instead, relatively broad ($\sigma_{Ne} \simeq 0.035$ 
keV and $\sigma_{Mg} \simeq 0.015$ keV, respectively), corresponding to velocity dispersion 
of $\sim 4\%$ and $\sim 1\%$ of the velocity of light, respectively. 
If our identification of the first two absorption lines is correct, than their energies
appear to be blue-shifted with respect the corresponding rest-frame energies of 
$\sim 0.025$ keV and $\sim 0.02$ keV, respectively, corresponding to a velocity of
$\sim 2.7 \%$ and $\sim 1.5 \%$ of the speed of light, possibly indicating the presence
of an outflowing, weakly relativistic wind. Note that, if the iron edge, that we detect at
$7.39$ keV, is indeed a blue-shifted neutral iron edge (rest-frame energy 7.112 keV), than
it would correspond to a velocity of $\sim 3.9 \%\, c$.

The addition of these components improved the quality of the
fit, returning a $\chi^2/dof = 3186/2759$, implying a decrease by $\Delta \chi^2 = 229$
for the addition of 12 parameters with respect to the previous best-fit. In this last 
fit, we decided to fix the outer disk radius at 1000 R$_g$ and the iron abundance at
2 times the Solar value, and to include the low-energy disklines also in the fitting 
of the {\nus} spectrum. The line at $2.6$ keV is indeed below the energy band used
for the {\nus} spectrum and therefore it was not included. The other two disklines
were included with all the parameters fixed to those of the {\xmm} spectrum, except for
the normalization that was left free to vary. We find that the line at 3.3 keV is indeed
not necessary in the {\nus} spectrum, with an upper limits on its equivalent width
of $30$ keV, while the addition of the line at $4.1$ keV in the {\nus} spectrum
is significant at $\sim 3 \sigma$ confidence level. All the other parameters are 
very similar (compatible within the associated errors) to those of the previous 
best-fit; the most significant difference is in the ionization parameter of the 
{\nus} spectrum for which we only get an upper limit of $\log \xi < 2$, in agreement 
with the centroid energy of the iron line at $\sim 6.4$ keV.
The results of this fit are shown in Table \ref{Tabfin}, and the residuals with
respect to the best fit model are shown in the bottom panel of Fig.\ \ref{fig4}.

\begin{table}
\caption{The best fit parameters of the spectral fitting of the {\xmm} ($0.6-10$
keV energy band) and {\xmm} + {\nus} ($0.6-70$ keV energy band) spectra of {\saxj}. 
The continuum emission is described by a combination of 
a blackbody and the Comptonization component {\it nthComp} included in the 
self-consistent reflection model {\it relxillCp}, modified at
lower energy by photoelectric absorption from neutral matter modeled with
{\it phabs}. The blackbody luminosity is given in units of $L_{36}/D_{10}^2$,
where $L_{36}$ is the bolometric luminosity in units of $10^{36}$~ergs/s and 
$D_{10}$ the distance to the source in units of 10~kpc. The blackbody radius
is calculated in the hypothesis of spherical emission and for a distance of
3.5~kpc. 
Uncertainties are given at $90\%$ confidence level.}
\begin{tabular}{|llcc|}
\hline
             &           & \multicolumn{2}{|c|}{RELXILLCP}   \\
  Component    & Parameter &  {\xmm} & {\nus} \\
             &           & \multicolumn{2}{|c|}{($0.6-70$ keV)} \\
\hline
phabs & $N_H$ ($\times 10^{22}$ cm$^{-2}$) &  \multicolumn{2}{|c|}{$0.125 \pm 0.011$}  \\

edge  & $E_{Fe}$ (keV)                &  \multicolumn{2}{|c|}{$7.39 \pm 0.05$} \\
edge  & $\tau$	($\times 10^{-2}$)	&  \multicolumn{2}{|c|}{$3.3 \pm 0.6$} \\

gauss & $E_{line}$ (keV)              &  \multicolumn{2}{|c|}{$0.947 \pm 0.006$} \\
gauss & $\sigma$ (keV)			&  \multicolumn{2}{|c|}{$0.035 \pm 0.007$}  \\
gauss & $I_{line}$ ($\times 10^{-4}$ ph cm$^{-2}$ s$^{-1}$)  &   \multicolumn{2}{|c|}{$-18.4 \pm 3.5$}  \\
gauss & $EqW$ (eV)	               & \multicolumn{2}{|c|}{$5.0 \pm 1.1$}   \\   
gauss & $E_{line}$ (keV)              &  \multicolumn{2}{|c|}{$1.372 \pm 0.006$} \\
gauss & $\sigma$ (keV)			&  \multicolumn{2}{|c|}{$0.015 \pm 0.006$}  \\
gauss & $I_{line}$ ($\times 10^{-4}$ ph cm$^{-2}$ s$^{-1}$)  &   \multicolumn{2}{|c|}{$-4.2 \pm 1.2$}  \\
gauss & $EqW$ (eV)	               & \multicolumn{2}{|c|}{$2.3 \pm 0.7$}   \\   
gauss & $E_{line}$ (keV)              &  \multicolumn{2}{|c|}{$1.596 \pm 0.014$} \\
gauss & $\sigma$ (keV)			&  \multicolumn{2}{|c|}{$0.03 \pm 0.02$}  \\
gauss & $I_{line}$ ($\times 10^{-4}$ ph cm$^{-2}$ s$^{-1}$)  &   \multicolumn{2}{|c|}{$-4.8 \pm 2.3$}  \\
gauss & $EqW$ (eV)	               & \multicolumn{2}{|c|}{$4.3 \pm 1.6$}   \\   

diskline & $E_{line}$ (keV)	      &  \multicolumn{2}{|c|}{$2.632 \pm 0.017$} \\
diskline & $I_{line}$ ($\times 10^{-4}$ ph cm$^{-2}$ s$^{-1}$)  & $32 \pm 6$  & $-$  \\
diskline & $EqW$ (eV)	               &  $72 \pm 12$  & $-$  \\      
diskline & $E_{line}$ (keV)	        &  \multicolumn{2}{|c|}{$3.34 \pm 0.02$} \\
diskline & $I_{line}$ ($\times 10^{-4}$ ph cm$^{-2}$ s$^{-1}$)  &  $21.9 \pm 0.3$  & $<22$ \\
diskline & $EqW$ (eV)	               &  $71 \pm 14$  & $<29$ \\       
diskline & $E_{line}$ (keV)	        & \multicolumn{2}{|c|}{$4.13 \pm 0.04$}  \\
diskline & $I_{line}$ ($\times 10^{-4}$ ph cm$^{-2}$ s$^{-1}$)  & $10.8 \pm 1.8$  &  $5.8^{+5.2}_{-2.8}$ \\
diskline & $EqW$ (eV)	              &  $48 \pm 10$  &  $37 \pm 19$ \\      
Smearing & $Betor$                   &  \multicolumn{2}{|c|}{$-2.30 \pm 0.04$}    \\  
Smearing & $R_{in}$ ($G M / c^2$)       &  \multicolumn{2}{|c|}{$<8.2$}   \\  
Smearing & $R_{out}$ ($G M / c^2$)  & \multicolumn{2}{|c|}{$1000$ (fixed)}   \\ 
Smearing & Incl (deg)               & \multicolumn{2}{|c|}{$68 \pm 4$} \\  

bbody & $kT_{BB}$ (keV)               & $0.111 \pm 0.004$ & $0.50^{+0.07}_{-0.33}$ \\
bbody & L$_{BB}$ ($L_{36}/D_{10}^2$)  & $2.7 \pm 0.8$ & $1.6^{+10}_{-0.3}$ \\
bbody & R$_{BB}$ (km)                 &  $130 \pm 21$ & $4.9^{+110}_{-1.5}$ \\   
nthComp& $\Gamma$                     & $1.77^{+0.02}_{-0.04} \pm 0.04$ & $1.939^{+0.008}_{-0.005}$  \\
nthComp & $kT_e$ (keV)		      & $7.3 \pm 0.8$ & $69^{+25}_{-15}$   \\
RelxillCp & Refl Frac               &  $0.78^{+0.7}_{-0.2}$  &  $0.35 \pm 0.06$  \\  
RelxillCp & Fe abund                &  \multicolumn{2}{|c|}{$2$ (fixed)}   \\  
RelxillCp & $\log \xi$              &  $3.64 \pm 0.12$  &  $<2$  \\  
RelxillCp & Norm ($\times 10^{-3}$) &  $2.7 \pm 0.2$   &   $4.14^{+0.10}_{-0.09}$  \\  
\hline
total &  $\chi^2$ (dof)	   & \multicolumn{2}{|c|}{$3186.45~(2759)$}   \\
\hline \\
\end{tabular}
\label{Tabfin}
\end{table}

\section{Discussion}

In this paper we have analyzed broad-band X-ray spectra acquired during the 2015
outburst of the AMSP {\saxj} observed by {\xmm} and {\nus}. The {\xmm} ToO was
performed at the peak of the outburst on 2015 April 11 for a total observing 
time of 110 ks, which resulted in an effective on-source exposure of $\sim 80$ ks.
The {\nus} observation   
was performed approximately four days later, on 2015 April 15, and resulted in
49 ks of exposure per each of the {\nus} modules. In this way, we have 
obtained a broad-band (from 0.6 to 70 keV), moderately high-resolution 
spectrum of the source. 

\subsection{Comparison with previous spectral results}

{\saxj} has been previously observed with good energy resolution by {\xmm} and 
{\it Suzaku}, approximately 1 day apart, during the 2008 outburst.
In that occasion, a broad iron line was detected in both the {\xmm} and 
{\it Suzaku} spectra, with very similar profiles, and was fitted by a {\it diskline}
\citep[see P09,][]{Cackett.etal:09}. The inner disk radius derived in this 
way was R$_{in} = 8.7 \pm 0.4$ R$_g$ using {\xmm} (P09) and R$_{in} = 13.2 \pm 2.5$ R$_g$
from a joint fit of the {\xmm} and {\it Suzaku} spectra \citep[][]{Cackett.etal:09},
respectively. In both cases, the inner disk radius was consistent with being
inside the co-rotation radius, which is the radius at which the magnetosphere 
rotation velocity equals that of an assumed Keplerian disc, R$_{co} = (G M_{NS}
/ \Omega^2)^{1/3}$, where $M_{NS}$ is the NS mass and $\Omega$ its spin angular
velocity. For the case of {\saxj}, the co-rotation radius is R$_{co} = 31\, 
m_{1.4}^{1/3}$ km, where $m_{1.4}$ is the NS mass in units of $1.4\, M_\odot$. 
This has to be compared with the inner disk radius inferred from the reflection
component, that is R$_{in} < 25.6\, m_{1.4}$ km (at $90\%$ confidence level, P09).
This is thought to be a necessary condition in order to observe coherent 
pulsations in accreting pulsars and to avoid efficient propeller ejection of 
matter due to the centrifugal barrier \citep[according to the standard theory
of accretion onto fast rotators, see e.g.][]{Ghosh.Lamb:1979}.
Assuming that the inner disk (as measured by the Fe line) is truncated at the 
magnetospheric radius this implies a magnetic field strength of $\sim 3 \times 10^8$
Gauss at the magnetic poles \citep{Cackett.etal:09}. Interestingly, this estimate
is consistent with other independent estimates of the magnetic field strength 
in this source based on completely different arguments \citep[see e.g.][and references therein]
{DiSalvo.Burderi:2003, Burderi.etal:2003, Burderi.etal:2006, Sanna.etal:2017b}.
Also, both the {\xmm} and {\it Suzaku} spectra gave a low ionization state
of iron, inferred from the line centroid always consistent with 6.4 keV
(corresponding to the rest-frame energy of the K$\alpha$ transition of
neutral or moderately ionized iron), and a high inclination angle of the 
system with respect to the line of sight, always above $50^\circ$.

Our analysis of the {\xmm} and {\nus} spectra of {\saxj} during the 2015
outburst gave remarkably similar results as regards the smearing parameters, 
both when fitting the iron line profile with a {\it diskline} and when 
using a self-consistent reflection model. These results are also very similar
to those obtained for the 2008 {\xmm} observation (P09).
To show the agreement between these results we report in Table 
\ref{tab3} the results obtained by P09 from the {\xmm} observation performed 
in 2008 and our results from the {\nus} spectrum obtained in 2015 when
fitting the line profile with a {\it diskline}. Again we find a low 
ionization parameter, a large inclination angle ($i > 70^\circ$), and 
a small inner disk radius (less than 7 R$_g$, corresponding to 
R$_{in} < 15\, m_{1.4}$ km). Even when we fit together the {\xmm} and
{\nus} spectra with a self-consistent reflection model, we get disk
parameters very similar to those obtained with the simple {\it diskline}.
In particular for the {\nus} spectrum, the ionization parameter, $\log \xi$, 
is less than 2.7, the emissivity index of the disk is around 2 (compatible 
with the presence of a central illuminating source), the inclination angle
is quite high, $> 50$ deg, and the inner disk radius is constrained 
between (best estimate) 7.5 and 11.5 R$_g$, corresponding to 
$16-24$ km for a 1.4 M$_\odot$ NS, well within the co-rotation radius. 
Note that, despite the fact that the spectrum is typical of a hard state, 
the inner disk radius is quite close to the NS surface, implying that the 
disk is truncated not too far from the compact object in the hard state. 
This is also observed in other NS/LMXBs in the hard state \citep[see]
[and references therein]{DiSalvo.etal:2015} and, as noted above, is a 
necessary condition to avoid a strong propeller effect in the case of
a pulsar.

\begin{table}
\caption{Comparison of the best-fit {\it diskline} parameters obtained
for the 2008 outburst as observed in the 60-ks {\xmm} observation 
\citep{Papitto.etal:09} and for the 2015 outburst as observed by {\nus} 
(this paper). For comparison we also show the best-fit {\it diskline} 
parameters obtained for the 2015 80-ks {\xmm} observation, when the source
appears to be in a soft state. 
Bol $L_X$ is the bolometric luminosity extrapolated in the 
$0.05 - 150$ keV energy range during the observation and assuming a distance
to the source of $3.5$ kpc.
}
\vskip 0.5cm
\begin{tabular}{|lccc|}
\hline
Parameter & {\xmm} (2008) & {\nus} (2015) & {\xmm} (2015)  \\
\hline
$E_{Fe}$ (keV)            & $6.43 \pm 0.08$  &  $6.38 \pm 0.10$  &  $6.70 \pm 0.05$ \\
$Betor$                   & $-2.3 \pm 0.3$ & $-2.0 \pm 0.2$  &   $-2.16 \pm 0.10$   \\  
$R_{in}$ ($G M / c^2$)    & $8.7 \pm 0.4$ & $< 7$   &   $10^{+8}_{-3}$  \\  
$R_{out}$ ($G M / c^2$)   & $127 \div 318$ & $280 \div 1070$  &  $760 \div 1780$  \\  
Incl (deg)                & $> 58$ & $> 70$   &  $>58$   \\
EqW (eV)                  & $120 \pm 20$  &  $120 \pm 20$  &   $58 \pm 8$   \\
Bol $L_X$ (erg/s) 	   & $6.6 \times 10^{36}$  &  $3.6 \times 10^{36}$  & $3.1 \times 10^{36}$ \\
\hline
\end{tabular}
\label{tab3}
\end{table}

A small inner disk radius is also implied for most of the other AMSPs 
for which a spectral analysis has been performed and a broad iron line 
has been detected in moderately high resolution spectra. 
The AMSP IGR J17511-3057, observed by {\xmm} for 70 ks and {\it RXTE} 
\citep{Papitto.etal:10}, shows both a broad iron line and the Compton hump
at $\sim 30$ keV. In this case, the inner disc radius was at $\ge 40$ km 
for a $1.4\, M_\odot$ NS, with an inclination angle between $38^\circ$ 
and $68^\circ$ \citep[see also][]{Papitto.etal:2016}. 
The AMSP and transitional pulsar IGR J18245-2452 observed
by {\xmm} \citep{Papitto.etal:2013a}, showed a broad iron line at 6.7 keV
(identified as K$\alpha$ emission from Fe XXV) with a width of $\sim 1.6$
keV, corresponding to R$_{in} \simeq 17.5$ R$_g$ or $\sim 36.7$ km for a 
$1.4\, M_\odot$ NS. For comparison, the inner disk radius derived from 
the blackbody component was $28 \pm 5$ km. The (intermittent) AMSP 
HETE J1900-2455, observed by {\xmm} for $\sim 65$ ks \citep{Papitto.etal:13b},
showed a broad iron line at 6.6 keV (Fe XXIII-XXV) and an intense and
broad line at $\sim 0.98$ keV, visible both in the pn and in the RGS
spectrum, compatible with being produced in the same disk region. 
In this case, the inner disc radius was $25 \pm 15\, R_g$, with an 
inclination angle between $27^\circ$ and $34^\circ$.
The (intermittent) AMSP SAX J1748.9-2021, observed by {\xmm} for $\sim 115$ ks 
and {\it INTEGRAL} \citep{Pintore.etal:2016}, was caught at
a relatively high luminosity of $\sim 5 \times 10^{37}$ erg/s corresponding
to $\sim 25\%$ of the Eddington limit for a $1.4\, M_\odot$ NS, and,
exceptionally for an AMSP, showed a spectrum compatible with a soft
state. 
The broad-band spectrum is in fact dominated by a cold thermal 
Comptonization component ($\sim 2$ keV) and shows an 
additional hard X-ray emission described by a power-law (photon index 
$\Gamma \sim 2.3$), typically detected in LMXBs in the soft state
\citep[see e.g.][]{DiSalvo.etal:2000}. In addition, a number of broad 
(Gaussian $\sigma = 0.1 - 0.4$ keV) emission features, likely associated 
to reflection processes, have been observed in the XMM-Newton spectrum.
A broad iron line was observed at an energy of $\sim 6.7-6.8$ keV, 
consistent with a Fe XXV K$\alpha$ transition produced in the disc at a 
distance of $\sim 20-43\, R_g$ ($\sim 42 - 90$ km) with an inclination 
angle of $\sim 38-45^\circ$. The other broad emission lines may be 
associated to K-shell emission of highly ionized elements, and are 
compatible with coming from the same emission region as the iron line. 
A moderately broad, neutral Fe emission line has been observed 
during the 2015 outburst of IGR J00291+5934 observed by {\xmm} and {\nus} 
\citep{Sanna.etal:2016}. Fitted with a Gaussian profile the line centroid 
was at an energy of $6.37 \pm 0.04$ keV with a $\sigma = 80 \pm 70$ eV,
while using a {\it diskline} profile, the line parameters were poorly
constrained. Finally, the newly discovered AMSP MAXI J0911-655, observed
by {\xmm} and {\nus} \citep{Sanna.etal:2016b}, shows the presence of a 
weak, marginally significant and relatively narrow emission line in the 
range $6.5-6.6$ keV, modelled with a gaussian profile with $\sigma$ ranging 
between 0.02 and 0.2 keV, which was identified with K$\alpha$ emission from 
moderate to highly ionized iron.

\subsection{Detailed discussion of the {\nus} and {\xmm} spectra: similarities and 
differences}

The use of a self-consistent reflection model instead of a {\it diskline} 
for the {\nus} spectrum of {\saxj} gives an improvement of the fit 
(corresponding to a $\Delta \chi^2 \simeq 9$ with one parameter less). 
This demonstrate that the Compton hump is significantly detected 
in the spectrum and that the line profile parameters are in good agreement 
with those needed to fit the whole reflection component.
The best-fit results strongly suggest a moderate overabundance of iron by 
approximately a factor between 2 and 3. However, this overabundance may be 
indicative of a disk density higher than the value of $10^{15}$ cm$^{-3}$ 
assumed in the {\it relxill} model. 
Indeed, \citet{Garcia.etal:2016} have shown that the iron abundance is sensitive 
to the density used in calculating the reflection model, and that when assuming 
a higher-density disk a lower iron abundance is obtained
\citep[see also][and references therein]{Garcia.etal:2018}. 
Hence, this overabundance should be confirmed using appropriate reflection
models in which the density in the disk can be varied. There are versions 
of the {\it relxill} model which allow to vary the disk density, although
at the moment these models have a high energy cutoff of the illuminating 
continuum fixed to 300 keV, that is much higher than the temperature of the
Comptonization continuum we find in the spectrum of {\saxj}.

We have also fitted the value of the outer radius of the emitting region 
in the disk; this parameter should be always let free to vary in high 
statistics spectra, since it strongly correlates with the inclination 
angle of the system and fixing this parameter may result in an artificially 
narrow uncertainty for the inclination angle. 
In all our fitting, the value of the outer radius of the emitting region in 
the disk, was quite undetermined, although compatible (within the large 
uncertainties) with the value derived by P09 during the 2008 outburst 
($130 \div 320\, R_g$).
Usually in other bright LMXBs, when it is possible to let this parameter free, 
the best fit value is most of the times very high, above $2500\, R_g$ 
\citep[see e.g.][]{Di_Salvo.etal:09, Iaria.etal:2009}, as it should be expected 
if the entire disk is illuminated and emits a reprocessed spectrum. However, {\saxj} 
is a transient with short outbursts, and therefore during the outburst at least 
the innermost part of the disk is emptying in few days. This means that the disc, 
over which reflection takes place, may have a ring-like shape with an outer disk 
radius which is relatively close to the inner disc radius. Therefore, it is 
possible that the small outer disk radius inferred from the reflection component 
in {\saxj} during the 2008 outburst may be an indication 
that the accretion disk was already emptying or that some disk parameters may 
change abruptly towards the outer disk in short-duration outbursts.
Unfortunately, we were not able confirm this result with these observations.
Future observations at high energy resolution and high statistics (perhaps taken
during the decay phase of the outburst) might be able to confirm this finding 
which may give important information on the evolution of the accretion disk in 
this kind of transients.

The best-fit continuum model that we find for the {\nus} spectrum of {\saxj} 
is also very similar to that already used by P09 to fit the {\xmm} spectrum 
during the 2008 outburst, with the difference that P09 used two soft components, 
a blackbody (at a temperature kT $\sim 0.4$ keV) and a multicolor disk blackbody
(at kT$_{in} \sim 0.2$ keV), and a power-law to fit the Comptonization
component, while we use a blackbody and a Comptonization model that include a soft 
(Wien spectrum) component as a seed photon. In our fit, the seed photon temperature 
for Comptonization (kT$_{seed} < 0.29$ keV) is   
comparable to the disk blackbody temperature in the P09 deconvolution, 
possibly indicating that both the disk and the NS contribute to the seed photons 
for Comptonization. The 2015 {\nus} spectrum also requires a blackbody component
with a temperature of $\sim 0.7$ keV, slightly higher than the blackbody temperature 
reported by P09, and a spherical radius of the emitting region of $\sim 3$ km, 
a factor $\sim 2$ lower than the blackbody radius reported by P09.  
The value of the equivalent hydrogen column to the source is very precisely
determined by the {\xmm} spectrum, $N_H \simeq 0.15 \times 10^{22}$ cm$^{-2}$, 
and is slightly lower   
than that derived by P09 during the 2008 outburst.

On the other hand the {\xmm} spectrum of {\saxj} taken in 2015 looks quite different
from the 2015 {\nus} spectrum and the 2008 {\xmm} spectrum of the source. The blackbody
component is found at a low temperature, $\sim 0.1$ keV, very close to the seed photon
temperature ($kT_{seed} \simeq 0.15$ keV), and the corresponding radius of the blackbody 
emitting region, assuming a spherical geometry and not considering color corrections, 
results to be $150 - 200$ km, much larger than the inner radius inferred from reflection 
features (cf.\ Tab.\ \ref{Tabtot}). 
The Comptonization spectrum appears much softer, with an electron temperature
around $5-8$ keV. Moreover, several emission lines are observed in the {\xmm} spectrum,
identified with $K\alpha$ transitions of highly ionized (He-like or H-like) elements 
(S XVI $-$ rest-frame energy $2.623$ keV, Ar XVIII $-$ rest-frame energy $3.323$ keV, 
Ca XIX$-$XX $-$ rest-frame energy $3.902$ and $4.108$ keV, respectively, and Fe XXV $-$ 
rest-frame energy $6.70$ keV). 
The smearing parameters of these lines are compatible to be
the same, and appear very similar to what we find for the smearing parameters of
the {\nus} spectrum (emissivity index -$2.1 - 2.3$, inner radius $6 - 18$ 
R$_g$, outer radius $\sim 1000$ R$_g$, inclination $50 - 65$ deg). When we try
to fit the reflection component (Fe line and Compton hump) with the self-consistent 
reflection model {\it relxill} we find parameters very similar to those obtained for 
the {\nus} spectrum, except for a high value of the ionization parameter 
($\log \xi \sim 3.7$), in agreement with the high energies of the emission lines, 
which would require $\log \xi > 2$. 
Most probably the {\xmm} spectrum of {\saxj} taken in 2015 corresponds to a transition
spectrum, in line with the fact that the {\xmm} observation was taken at the very beginning 
of the outburst, that evolved to the more standard {\nus} spectrum a few days after.
If this is the case, {\saxj} experienced a soft to hard transition at the
beginning of the outburst, that has never been observed before for an AMSP. Note
that a spectral transition has been observed for the 11 Hz X-ray pulsar IGR J17480-2446
in the globular cluster Terzan 5 during its X-ray outburst in 2010 
\citep[][]{Papitto.etal:2012}. However, in that case, a hard to soft state transition was 
observed during the outburst rise. 
Unfortunately, the lack of high-energy coverage strictly simultaneous to the {\xmm} 
observation of {\saxj} does not allow us to put further constraints on the high-energy 
spectrum, or to look for the presence of hard continuum components or a complex 
reflection component.


\subsection{Binary inclination and mass of the neutron star}

As stated above, despite the differences between the {\xmm} and {\nus} spectra we 
find quite similar values for the parameters of the reflection component and relativistic
smearing. We therefore tried to fit simultaneously these spectra in order to
increase the statistics of the reflection features and improve the constraints
on the corresponding best-fit parameters. We therefore let free to vary all
the parameters of the continuum emission, except for the $N_H$, 
that was not constrained 
in the {\nus} spectrum alone, and tied together the parameters of the relativistic 
smearing, with the iron abundance fixed at 2 times the Solar value.
Fitting the {\xmm} and {\nus} spectra together, we could get a precise estimate of 
the inner disk radius, which is constrained between 7.5 and 11.5 $R_g$, while the
best estimate of the system inclination, constrained between $58^\circ$ and $64^\circ$, 
comes from the fitting with {\it relxillCp} of the {\xmm} spectrum (see Tab.\ \ref{Tabtot}). 

We also find evidence in the {\xmm} spectrum of the presence of some absorption 
discrete features (see Tab.\ \ref{Tabfin}), namely an absorption edge at $\sim 7.4$ keV 
from neutral or mildly ionized iron and at least two absorption lines, possibly from 
$K\alpha$ transitions of highly ionized (He-like) Ne IX (at 0.947 keV) and Mg XI 
(at 1.372 keV). 
These lines appear relatively broad (implying a velocity dispersion of $\sigma_v \sim 1 \% \,c$) 
and blue-shifted at a velocity a few percent the speed of light. If confirmed, these 
lines may suggest the presence of a weakly relativistic outflowing wind towards the observer.
Absorption lines from ionized elements are usually observed in high-inclination 
($60-70^\circ$) sources \citep[see e.g.][and references therein]{Pintore.etal:2014}
and therefore their presence in the {\xmm} spectrum may support the possibility of a 
high inclination, $i \sim 60^\circ$, in {\saxj}.

This estimate is compatible with previous estimates based on the fitting of the 
iron line profile during the 2008 outburst observed by {\xmm} \citep[P09,][]
{Cackett.etal:09}, and is also consistent with the inclination of $60^\circ 
\pm 5^\circ$ given by \citet{Ibragimov.Poutanen:2009} from a detailed analysis
of the 2002 outburst from the source, as well as with the inclination range from
$36^\circ$ to $67^\circ$ given by \citet{Deloye.etal:2008} studying the 
optical modulation along the orbital period of the system observed during a 
quiescence period in 2007; these authors also suggest a pulsar mass $> 2.2\, 
M_\odot$. A high inclination is qualitatively in agreement with the claim 
of a massive NS ($> 1.8\, M_\odot$) and a low mass companion star, 
a brown dwarf with $< 0.1\, M_\odot$, as suggested by \citet{Bildsten.etal:2001}
\citep[see also][]{DiSalvo.etal:2008, Burderi.etal:2009}. Finally, a 
recent estimate of the inclination angle to the system comes from a time-resolved 
optical imaging of SAX J1808.4-3658 during its quiescent state and 2008 outburst.
A Markov chain Monte Carlo technique has been used to fit the multi-band light 
curve of the source in quiescence with an irradiated star model, and a tight 
constraint of $50^{+6}_{-5}$ deg has been derived on the inclination angle
\citep[][]{Wang.etal:2013}. This implies a constraint on the mass of the pulsar 
and its companion star, which are inferred to be $0.97^{+0.31}_{-0.22}\, M_\odot$
and $0.04^{+0.02}_{-0.01}\, M_\odot$ (both at $1\sigma$ confidence level), 
respectively.

However, high values for the inclination angle of the system look at odd 
when considered together with optical estimates of the radial velocity of
the companion star. From phase resolved optical spectroscopy and photometry 
of the optical counterpart to {\saxj}, obtained during the 2008 outburst,
\citet{Elbert.etal:2009} reveals a focused spot of emission at a location 
consistent with the secondary star. The velocity of this emission is estimated
at $324 \pm 15$ km/s; applying a “K-correction”, the authors estimate the velocity 
of the secondary star projected on to the line of sight to be $370 \pm 40$ km/s
\citep[see also][]{Cornelisse.etal:2009}. This estimate, coupled with a 
high inclination angle of the system, gives very low values for the NS mass, 
and has been used to argue against the presence of a heavy NS in this system.
In fact, the pulsar mass can be estimated using the following relation: 
$M_1 \sin^3 i / (1+q)^2 =  K_2^3 P_{orb}/(2\pi G)$, where $M_1$ is the pulsar
mass, $q = M_2/M_1$ is the mass ratio of the system, $P_{orb}$ is the orbital
period of the system, and $K_2$ is the radial velocity of the companion star
of mass $M_2$. Using the estimated radial velocity of the companion star 
together with our best-fit value for the inclination angle, we find a
pulsar mass in the range: $M_1 = 0.5 \div 0.8 \, M_\odot$. This 
range of masses for a NS is unacceptable and casts serious doubts on the 
estimates of the radial velocity of the companion and/or on a high inclination
angle for the system. 

A possibility we can imagine is that the reflection is measuring the inclination
with respect to the sight of the inner part of the accretion disk, that may be 
different from the binary inclination. If the inner accretion disk is tilted with 
respect to the orbital plane, for instance because of the action of the NS magnetic 
field, such that the inner disk is observed at high inclination, than this could 
explain why measured inclination of the inner disk can be different from the binary 
inclination. However, this would not explain the high inclination angle measured
by \citet[][]{Wang.etal:2013} during X-ray quiescence.

The other possibility is that
the problem comes from measurements of the companion radial velocity.
Note that the reported measurements of the radial velocity
$K_2$ are still affected by large uncertainties. This is because these measurements
are taken during X-ray outburst and are affected by the presence of the accretion
disk and the strong irradiation of the companion star. These estimates should 
therefore be confirmed in order to obtain a reliable estimate of the NS mass.

\section{Conclusions}

In summary, we have reported a detailed spectral analysis of the {\xmm} and {\nus}
spectra of {\saxj} during the latest outburst in 2015. The main results of this 
study are described in the following.
The {\xmm} spectrum, taken for the first time at the beginning of the outburst, 
appears to be much softer than what is usually found for this source and quite
puzzling, while the broad-band {\nus} spectrum, acquired a few days after, gives 
results perfectly compatible with those found from the {\xmm} observation 
performed in 2008.
Despite the differences present between the {\xmm} and {\nus} spectra taken
in 2015, we could fit simultaneously the smeared reflection component in these
spectra. In particular, we find that the reflection component requires a ionization 
parameter of $\log \xi \sim 2.4$ for the {\nus} spectrum and a higher value,
$\log \xi \sim 3.8$ for the {\xmm} spectrum, 
and strong evidence of an overabundance of iron by a factor two with respect to 
the solar abundance, although this may be due to a relatively high density in the disk.
Also, the smearing parameters are very similar to those found with {\xmm} 
during the 2008 outburst. 
The emissivity index of the disc is $\sim -2$, consistent with a dominating 
illuminating central source, 
and we find that the upper limit to the inner disk radius is $\sim 12~R_g$,
compatible with an inner disk radius smaller than the corotation radius.
We also give a precise measure of the inclination angle of the system,
which results around $60^\circ$, in agreement with previous spectral results,
as well as with the results of fitting the reflection component in each spectrum 
with empirical models (disklines). A high-inclination angle for this system is
also supported by the presence of absorption discrete features in the {\xmm} 
spectrum, although these detections should be confirmed by further spectroscopic 
studies.
The high inclination of the system with respect to our line of sight, when 
combined with available measurements of the radial velocity of the optical 
companion, poses, however, a problem as regards the correct determination of 
the mass of the NS in this systems, and therefore deserves further investigation.

\thanks{We thank the unknown referee for her/his suggestions that certainly
improved the quality of the manuscript. We also thank Dr.\ Matranga for collaborating
to a first draft of this paper. 
We acknowledge financial contribution from the agreement ASI-INAF I/037/12/0.
A.P. acknowledges funding from the European Union’s Horizon 2020
Framework Programme for Research and Innovation under the Marie Sklodowska-Curie
Individual Fellowship grant agreement 660657-TMSP-H2020-MSCA-IF-2014.
We acknowledges support from the HERMES Project, financed by the Italian Space Agency 
(ASI) Agreement n.\ 2016/13 U.O, as well as fruitful discussion with the international 
team on “The disk-magnetosphere interaction around transitional millisecond pulsars” 
at the International Space Science Institute, Bern.
}

\bibliography{ms_saxj_NuSTAR}

\end{document}